\documentclass{article}
\usepackage{graphicx} 
\usepackage{underscore} 
\usepackage{url} 
\usepackage{hyperref} 
\usepackage{lineno}

\usepackage{geometry}
\usepackage{lineno}

\usepackage{sectsty}
\sectionfont{\fontsize{12}{12}\selectfont}
\subsectionfont{\fontsize{11}{11}\selectfont}

\usepackage{setspace}
\usepackage{titlesec}
    \titlespacing{\section}{0pt}{\parskip}{2pt}
    \titlespacing{\subsection}{0pt}{\parskip}{2pt}
    \titlespacing{\subsubsection}{0pt}{\parskip}{2pt}
    \titlespacing{\paragraph}{0pt}{\parskip}{6pt}

\usepackage[utf8]{inputenc}
\usepackage{csquotes}
\usepackage{amsmath,amssymb,mathtools}
\usepackage{enumitem}
\usepackage[normalem]{ulem}

\usepackage{placeins}
\usepackage{tikz}
\usepackage{multirow}
\usepackage{booktabs} 
\usepackage[flushleft]{threeparttable} 
\usepackage{graphicx}
\usepackage[]{caption}
\usepackage{wrapfig}
\usepackage{floatrow}
\usepackage{xcolor}

\usepackage{hyperref} 
\hypersetup{
	colorlinks,
	citecolor=blue,
	filecolor=blue,
	linkcolor=blue,
	urlcolor=blue
}
\definecolor{TitlePurple}{RGB}{102, 102, 255}
\definecolor{TitleBlue}{RGB}{4, 4, 204}
\definecolor{AssignColor}{RGB}{212, 4, 49}
\definecolor{BoxBack}{RGB}{222, 222, 222}
\definecolor{BoxTitle}{RGB}{204, 2, 2}

\usepackage[backend=bibtex,style=nature,natbib=true,url=false,date=year,doi=true,isbn=false,eprint=false]{biblatex}
\AtEveryBibitem{\clearlist{language}}

\definecolor{lime}{HTML}{A6CE39}
\DeclareRobustCommand{\orcidicon}{%
	\begin{tikzpicture}
	\draw[lime, fill=lime] (0,0) 
	circle [radius=0.16] 
	node[white] {{\fontfamily{qag}\selectfont \tiny ID}};
	\draw[white, fill=white] (-0.0625,0.095) 
	circle [radius=0.007];
	\end{tikzpicture}
	\hspace{-2mm}
}

\foreach \x in {A, ..., Z}{%
	\expandafter\xdef\csname orcid\x\endcsname{\noexpand\href{https://orcid.org/\csname orcidauthor\x\endcsname}{\noexpand\orcidicon}}
}

\usepackage[affil-it]{authblk}

\title{Interactively visualizing biological multilayer networks using MiRA}

\author[1]{Shir Miryam Nehoray\orcidB{}}
\author[1]{Yuval Bloch\orcidC{}}
\author[1,*]{Shai Pilosof\orcidA{}}
\affil[1]{Department of Life Sciences, Ben-Gurion University of the Negev, Beer-Sheva, Israel}
\affil[*]{Corresponding author: pilos@bgu.ac.il}

\usepackage{newfloat}
\usepackage{subcaption}

\begin{document}

\maketitle

\section{Abstract}

Multilayer networks are widely used across biology to represent systems in which complex networks vary across space, time, or interaction types. However, interactive visualization tools remain limited. We present MiRA (Multilayer Interactive Rendering Application), a browser-based, installation-free web application for visualizing biological multilayer networks. MiRA offers seven complementary visualization modes and interactive features that enable researchers to visually navigate the high complexity of multilayer networks for research and education.

\section{Main}

Biological systems are inherently complex. Neurons form circuits, genes regulate one another, and species interact across trophic levels. In each case, understanding the system requires mapping the web of interacting components. However, biological systems cannot be fully captured by a single network. Ecosystems change over time and space; the same genes are co-expressed across tissues; brain networks vary across pathological conditions; and protein interactions are modulated by cellular context. Multilayer networks provide a principled framework for representing this reality, encoding multiple interaction types, spatial contexts, or temporal snapshots within a unified structure\cite{Kivela2014-bj}. Their adoption across ecology \cite{Pilosof2017-gq}, neuroscience \cite{Vaiana2020-sy}, epidemiology \cite{Sood2023-cg}, microbiology \cite{Shapiro2023-bj}, and systems biology \cite{Hammoud2020-vg, Galai2023-io} reflects how fundamental this need is \cite{Aleta2026-up, De-Domenico2023-he}.

Despite the growing use of multilayer networks, interactive visualization tools have lagged far behind analytical methods \cite{McGee2019-yr}. This is a significant problem because visualization is often the first and most essential step in making sense of complex data. Designing visualization tools for multilayer networks is genuinely difficult, as such tools must simultaneously convey within-layer topology, interlayer connectivity, node identity across layers, and, where relevant, geographic or temporal context. The only two existing dedicated tools leave critical gaps. MuxViz \cite{De-Domenico2015-mh,De-Domenico2022-sf} is the most established platform for multilayer network analysis but requires programming knowledge in R and produces visualizations that are not interactively explorable in a browser. Arena3Dweb \cite{Karatzas2021-yr,Kokoli2023-fu} offers 3D web-based visualization but was designed for biomedical networks and does not support bipartite structures, which are among the most common network types in ecology (e.g., plant--pollinator, host--parasite, and seed dispersal networks) and in systems biology (e.g., gene--protein interactions). Both of these lack support for geographical contexts.

Here, we present MiRA (Multilayer Interactive Rendering Application), a browser-based, installation-free web application. We followed guidelines for multilayer network visualization and designed it with biological systems in mind~\cite{McGee2019-yr}. Full  documentation of all features is maintained alongside the software and updated continuously. MiRA is freely accessible at \url{https://mira.ecomplab.com} and runs directly on any computer browser without installation. Multilayer networks can be uploaded via JSON or CSV files; MiRA is also tightly integrated with the EMLN R package~\cite{Frydman2023-hg}, where objects of class \texttt{multilayer} can be directly visualized with a single line of code. Visualization sessions can be saved and reloaded, and users can export print-quality screenshots for use in publications. MiRA is designed exclusively for visualization: multilayer network analysis is already well served by dedicated packages, and focusing on visualization allows us to deliver the best possible user experience for visual exploration. MiRA calculates several basic multilayer network properties (see Methods) to aid researchers in their first exploration of the network, before they apply any analyses themselves. These properties are also used internally by the visualization itself (e.g., for node sizing and color mapping). However, users are encouraged to perform their analyses upstream and store results (such as module affiliations or centrality scores) as node, link or layer attributes in the data uploaded to MiRA, where they become immediately available for color mapping and filtering.

\begin{figure}[h]
    \centering
    \includegraphics[width=1.0 \linewidth]{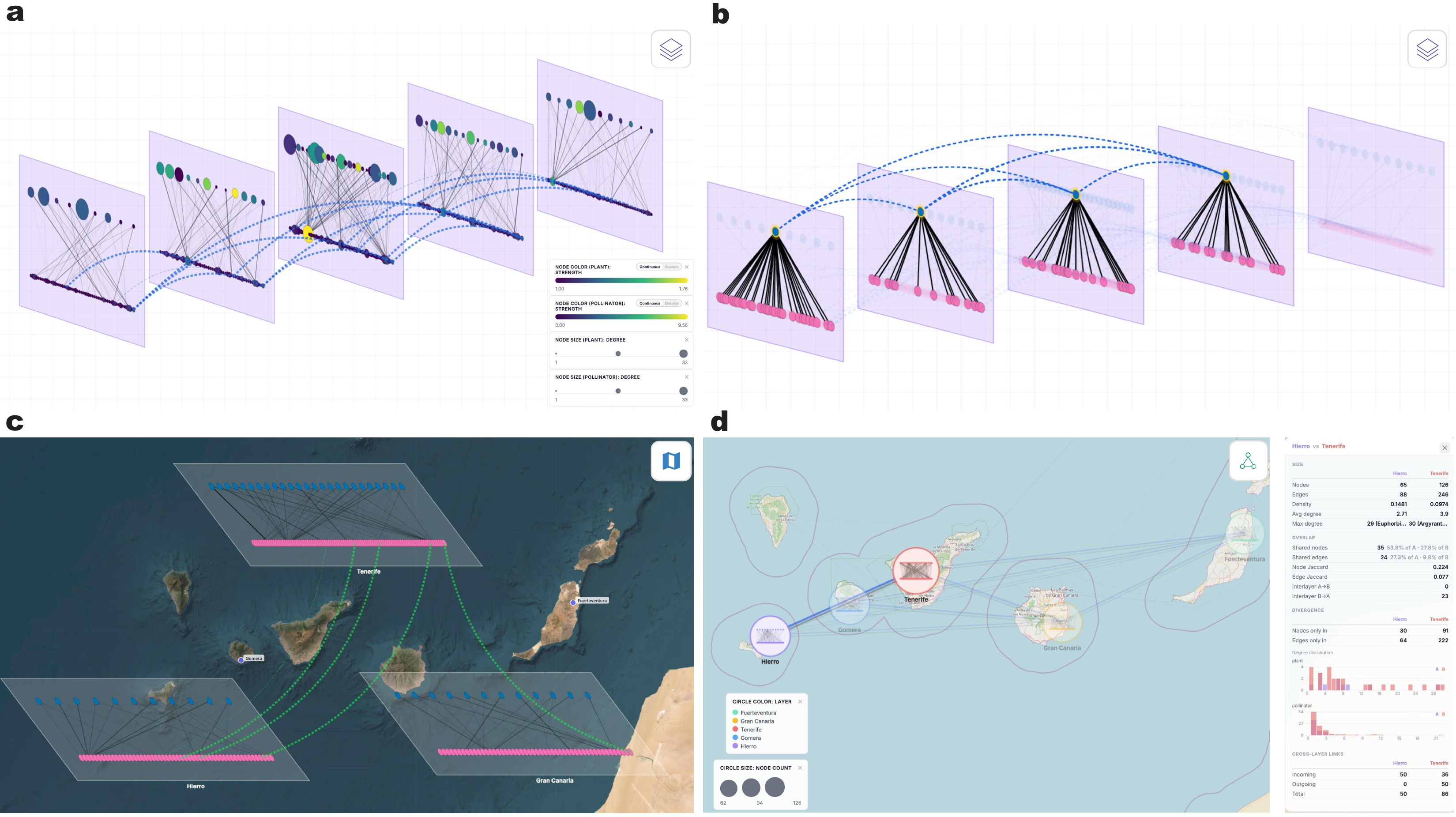}
    \caption{ \textbf{Network Mode, Map Mode, and Layer Mode visualization of the Canary Islands pollination multilayer network.} The icon in the top-right corner represents the active visualization mode. \textbf{(a)} Network Mode showing all layers simultaneously in a 3D stacked-layer view. Nodes are colored by strength and scaled by degree (corresponding legends are shown in the legend panels) for plants and pollinators separately. Interlayer links are displayed with a weight threshold of 0.55 for visual clarity. \textbf{(b)} Node selection: a selected plant species is highlighted across all layers in which it occurs, together with its intralayer and interlayer links while all other nodes are dimmed. \textbf{(c)} Map Mode with layers positioned geographically at their sampling locations. Blue markers indicate layer positions and three selected layers are extracted into floating 3D network panels. Layer panel and interlayer link colors are user-configurable; colors shown here were selected for visual clarity. \textbf{(d)} Layer Mode in Geographic Layout with a street map visualization. Gray edges represent shared nodes between layers and blue edges represent interlayer link counts. Two layers selected simultaneously (Cmd/Ctrl + click) and pairwise comparison panel activated (presenting on the right side of the map).}
    \label{fig:fig.1}
\end{figure}

\begin{figure}[h]
    \centering
    \includegraphics[width=1.0\linewidth]{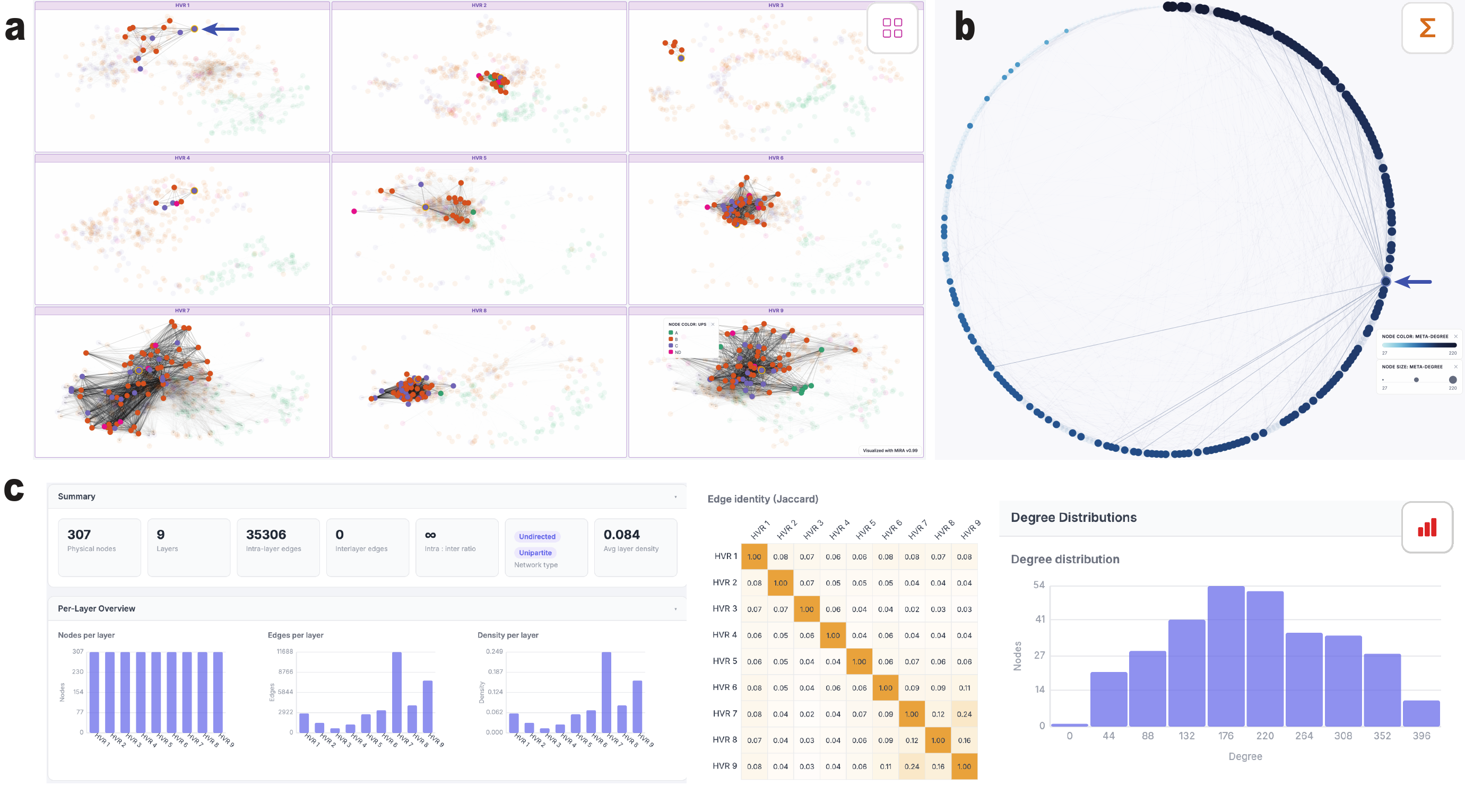}
    \caption{\textbf{Grid, Meta-network and Dashboard Mode visualization of the \textit{Plasmodium} \textit{var} gene multilayer network}. The icon in the top-right corner represents the active visualization mode. \textbf{(a)} Grid mode, showing all 9 layers. A selected node (indicated with an arrow) and its neighbors are highlighted, all other nodes and links are dimmed. \textbf{(b)} Meta-Network Mode showing the aggregated network across all layers, with edges combined by sum of weights. Links above the threshold value are shown, the rest are dimmed. Node selection is carried out across modesl: the same node that was selected in Grid mode is also highlighted. \textbf{(c)} Several examples of statistics shown in the Dashboard Mode panels. From left to right: the Summary panel; per layer distributions; a Jaccard similarity heatmap representing similarity in edge identity across layers and the degree distribution.}
    \label{fig:fig.2}
\end{figure}

Because multilayer networks carry inherently higher visual complexity than monolayer networks, interpretation at scale becomes considerably more difficult than in standard 2D visualization \cite{McGee2019-yr}. MiRA is designed to navigate this visual complexity using seven visualization modes. \textbf{(1) Network Mode} is the default view: an interactive 3D stacked-layer representation showing within-layer topology and interlayer connectivity simultaneously. Nodes can be colored according to network properties such as degree, strength or user-provided node attributes (e.g., abundance, module) (Fig.~\ref{fig:fig.1}a). Individual nodes can be isolated by clicking on a specific node or by using the node search sidebar; the selected node and its connections are then highlighted across all layers (Fig.~\ref{fig:fig.1}b). \textbf{(2) Map Mode} is available when the data uploaded include geographic coordinates, positioning each layer on an interactive map; individual layers can be popped out into floating 3D network panels for side-by-side spatial comparison (Fig.~\ref{fig:fig.1}c). \textbf{(3) Layer View Mode} renders layers themselves as the primary visual objects: each layer appears as a bubble with a micro-graph preview of its internal topology, positioned by a force-directed algorithm so that layers sharing many nodes or strong interlayer coupling are drawn close together. Selecting a second layer via Cmd/Ctrl+click opens a pairwise comparison panel showing shared nodes, Jaccard similarity, shared edges, and overlaid degree distributions. When every layer in the dataset has latitude and longitude, Layer View Mode activates in geographic layout automatically, whereby bubbles are pinned to their real-world coordinates on a map background (Fig.~\ref{fig:fig.1}d). \textbf{(4) Grid View Mode} renders each layer in its own mini panel arranged in a responsive grid. All cells share the same axes and node identities, making it easy to compare intralayer structure across layers side-by-side without the visual clutter of the 3D stack (Fig.~\ref{fig:fig.2}a). \textbf{(5) Meta-Network Mode} aggregates all intralayer links across layers into a single aggregated network, allowing cross-layer patterns to be viewed as a unified monolayer with user-controlled aggregation options. Selecting nodes and links provides information on their appearance across the layers (Fig.~\ref{fig:fig.2}b).

These five modes are designed to reveal structural patterns through interactive visual exploration. MiRA also provides two complementary modes for exploring the network. \textbf{(6) Dashboard Mode} provides a statistical overview, replacing the network canvas with descriptive statistical panels: per-layer node, edge, and density charts; intralayer and interlayer link weight distributions; a Participation section combining a node-by-layer presence matrix with a multiplexity histogram; Jaccard similarity heatmaps for node and edge composition across layers; and degree and strength distributions split into intralayer and interlayer components (further split into in/out panels for directed networks; Fig.~\ref{fig:fig.2}c). \textbf{(7) Data Mode} provides an interactive tabular interface for filtering and inspecting the raw data. It displays the full network as interactive, filterable tables of nodes, layers, edges, and state nodes, with real-time updates to the visualization as filters are applied.

Altogether, MiRA fills a critical gap in the biological network analysis toolkit. By allowing researchers to visualize multilayer data interactively in a browser, without installation or programming, it lowers the entry barrier for multilayer approaches across disciplines and research contexts. MiRA does not represent a mere incremental improvement but a qualitative shift in accessibility and capability: for the first time, researchers across all biological disciplines can interactively explore multilayer network data without sacrificing representational richness. The interactive interface allows researchers to engage with the inherent complexity of multilayer data in a natural and intuitive way---filtering, selecting, and switching between modes---rather than confronting it as static images or tabular summaries. Beyond research, these capabilities make MiRA a highly useful tool for education, as interactive visualization is one of the most powerful means of conveying complex network structure to students and broader audiences.

\section{Methods}

\subsection*{Internal network data model}
Multilayer networks have five components \cite{Kivela2014-bj}:

\begin{itemize}
    \item \textit{Layers}, which represent the state of the system (e.g., locations in space, temporal points, interaction types). In some networks, layers exist in more than one dimension; for example, when a network changes in space and in time. Due to the already high visual complexity of multilayer networks even in a single dimension, MiRA currently supports only a single dimension.
    \item Physical nodes, which are the physical entities in the network (e.g., genes, proteins, species);
    \item State nodes, which are the realization of a physical node in a given layer;
    \item Intralayer links, which encode the relationships between nodes within each layer;
    \item Interlayer links, which encode node connectivity across layers. Interlayer links are not always present.
\end{itemize}

This complexity can be encoded in computational data structures in multiple ways (e.g., \cite{De-Domenico2015-mh,Frydman2023-hg}). The internal data model of MiRA is based on the data structures defined by the EMLN R package \cite{Frydman2023-hg}, which provide a systematic data structure that encodes all the multilayer components in a single object. This design choice ensures that the data structure is well-tested and grounded in the mathematical foundations of multilayer network theory \cite{Kivela2014-bj,Pilosof2017-gq,Frydman2023-hg}. 

Network data are represented as a JSON object with four required arrays (\texttt{layers}, \texttt{nodes}, \texttt{extended}, \texttt{state\_nodes}) and two optional Boolean flags (\texttt{directed}, \texttt{directed_interlayer}) that control arrowhead rendering. The \texttt{directed} flag applies to intralayer edges and \texttt{directed_interlayer} to interlayer edges; setting \texttt{directed} to \texttt{true} forces \texttt{directed_interlayer} to \texttt{true} as well, and either flag can also be inferred from per-link \texttt{directed} properties on individual edges. The \texttt{layers} array defines each layer's identity, optional geographic coordinates (\texttt{latitude}, \texttt{longitude}), and bipartite status; any extra fields become selectable attributes for layer color mapping. The \texttt{nodes} array defines physical nodes, carrying a \texttt{node_type} field required for bipartite layers and any additional network-wide attributes. The \texttt{extended} array encodes both intralayer and interlayer edges in a single extended edge list with required fields \texttt{layer_from}, \texttt{node_from}, \texttt{layer_to}, and \texttt{node_to}, plus an optional \texttt{weight} that drives weight thresholding and link color mapping; intralayer and interlayer edges are distinguished automatically by whether \texttt{layer_from} equals \texttt{layer_to}. The \texttt{state_nodes} array enumerates every (layer, node) pair present in the network and carries per-layer node attributes whose values can differ for the same node across layers (e.g., \texttt{abundance} or \texttt{module}), as distinct from the network-wide attributes in the \texttt{nodes} array. This schema is the \textit{lingua franca} between all data entry paths (manual JSON upload, CSV import, and EMLN integration), and all visualization modes, ensuring that any valid input is immediately available to any mode without transformation.

\subsection*{Calculated network properties}

MiRA computes a small set of basic structural properties for the network it is rendering. These calculations serve two purposes: they provide quantitative summaries that researchers can inspect during their first exploration of a dataset (primarily through Dashboard Mode), and they feed the visualization itself --- for example, by mapping node degree or strength to node size and color in Network Mode, or producing the statistics for Dashboard Mode. By design, MiRA does not implement higher-order analyses (e.g., centrality, community detection, motif analysis, or null-model comparison): these are well served by dedicated multilayer analysis packages \cite{De-Domenico2015-mh,Farage2021-lg}, and users are encouraged to compute them upstream and store results as node, link, or layer attributes in the data uploaded to MiRA, where they become immediately available for color mapping and filtering.

\paragraph{Notation.}
Throughout, we follow the matrix conventions of Boccaletti et al.\ \cite{Boccaletti2014-vr}, with minor adaptations to support weighted networks. Let $L$ denote the number of layers and $\alpha,\beta\in\{1,\dots,L\}$ index layers. We write $X_{\alpha}$ for the set of physical nodes present in layer $\alpha$, $N_{\alpha}=|X_{\alpha}|$, and $E_{\alpha}$ for the set of (deduplicated) edges in layer $\alpha$. The intralayer adjacency matrix of layer $\alpha$ is $A^{[\alpha]}=(a^{\alpha}_{ij})$, where $a^{\alpha}_{ij}\geq 0$ is the weight of the edge from node $i$ to node $j$ in layer $\alpha$ ($a^{\alpha}_{ij}=0$ if absent; defaults to $1$ when the data does not specify a weight). The interlayer adjacency matrix between layers $\alpha\neq\beta$ is $A^{[\alpha,\beta]}=(a^{\alpha\beta}_{ij})$, defined analogously and including both diagonal ($j=i$, replica) and off-diagonal ($j\neq i$, general) couplings \cite{Kivela2014-bj,De-Domenico2013-xe}. The indicator $\mathbf{1}[\cdot]$ takes value $1$ when its argument is true and $0$ otherwise; intralayer self-loops are excluded from degree and strength via the $j\neq i$ convention.

Table~\ref{tab:mira-metrics} lists the core formulas. When the intralayer or interlayer edge type is directed, MiRA reports in/out variants of (1)--(4) by replacing $a^{\alpha}_{ij}$ with $a^{\alpha}_{ji}$ (incoming) or keeping $a^{\alpha}_{ij}$ (outgoing) in the obvious way; the undirected scalar is dropped when its directed counterparts are present. For bipartite layers, Jaccard similarity is additionally reported separately for each of the two node sets. Distributions shown in Dashboard Mode (degree, strength, weight, multiplexity histograms) are derived directly from the scalars in Table~\ref{tab:mira-metrics} by linear binning; the layer-by-layer presence matrix is the binary indicator $\mathbf{1}[v\in X_{\alpha}]$ underlying the participation count $P(v)$.

The meta-network projection used in Meta-Network Mode is a generalisation of the projected graph $\mathrm{proj}(\mathcal{M})=(X_{\mathcal{M}},E_{\mathcal{M}})$ of Boccaletti et al.\ \cite{Boccaletti2014-vr} (Eq.~4 therein), with three user-selectable choices for how layer-specific edge weights are aggregated onto the projected meta-edge (rows 15--17 of Table~\ref{tab:mira-metrics}).

\begin{table}[htbp]
\caption{\textbf{Multilayer network properties calculated by MiRA.} State-node-level metrics (1--4) are defined per (node, layer) pair $(i,\alpha)$. Per-physical-node aggregates (5--7) collapse them across the layers in which a node $v$ appears. Layer-level (8--12) and pairwise (13--14) metrics describe individual layers and their relationships. Meta-graph quantities (15--19) describe the layer-aggregated projection used in Meta-Network Mode. Notation follows Boccaletti et al.\ \cite{Boccaletti2014-vr}.}
\label{tab:mira-metrics}
\small
\begin{threeparttable}
\begin{tabular}{@{}p{0.4cm}p{4.2cm}p{5.7cm}p{5.0cm}@{}}
\toprule
\textbf{\#} & \textbf{Property} & \textbf{Formula} & \textbf{Notes} \\
\midrule
\multicolumn{4}{@{}l}{\textit{State-node level (per node $i$, per layer $\alpha$)}} \\
\addlinespace[2pt]
1 & Intralayer degree & $k^{\alpha}_{i,\,\mathrm{intra}} = \displaystyle\sum_{j\neq i} \mathbf{1}\!\left[a^{\alpha}_{ij}>0\right]$ & Self-loops excluded. Directed networks: in/out variants \\[6pt]
2 & Intralayer strength & $s^{\alpha}_{i,\,\mathrm{intra}} = \displaystyle\sum_{j\neq i} a^{\alpha}_{ij}$ & Weighted analogue of (1) \\[6pt]
3 & Interlayer degree & $k^{\alpha}_{i,\,\mathrm{inter}} = \displaystyle\sum_{\beta\neq\alpha}\sum_{j} \mathbf{1}\!\left[a^{\alpha\beta}_{ij}>0\right]$ & Inner sum over all $j$, including $j=i$ (replica couplings) \\[6pt]
4 & Interlayer strength & $s^{\alpha}_{i,\,\mathrm{inter}} = \displaystyle\sum_{\beta\neq\alpha}\sum_{j} a^{\alpha\beta}_{ij}$ & Weighted analogue of (3) \\
\addlinespace[3pt]
\midrule
\multicolumn{4}{@{}l}{\textit{Per-physical-node aggregates (over layers in which node $v$ appears)}} \\
\addlinespace[2pt]
5 & Layer-sum aggregate & $f_{\mathrm{sum}}(v) = \displaystyle\sum_{\alpha:\,v\in X_{\alpha}} f(v,\alpha)$ & $f\in\{k_{\mathrm{intra}},\,s_{\mathrm{intra}},\,k_{\mathrm{inter}},\,s_{\mathrm{inter}}\}$ \\[6pt]
6 & Layer-mean aggregate & $f_{\mathrm{mean}}(v) = f_{\mathrm{sum}}(v)\,/\,P(v)$ & Mean over layers in which $v$ is present \\[6pt]
7 & Participation (multiplexity) & $P(v) = \left|\{\alpha : v\in X_{\alpha}\}\right|$ & Number of layers in which $v$ appears \\
\addlinespace[3pt]
\midrule
\multicolumn{4}{@{}l}{\textit{Layer level (per layer $\alpha$)}} \\
\addlinespace[2pt]
8 & Density (unipartite, undirected) & $\rho(\alpha) = \dfrac{|E_{\alpha}|}{N_{\alpha}(N_{\alpha}-1)/2}$ & On deduplicated edge set; self-loops excluded \\[10pt]
9 & Density (unipartite, directed) & $\rho(\alpha) = \dfrac{|E_{\alpha}|}{N_{\alpha}(N_{\alpha}-1)}$ & \\[10pt]
10 & Density (bipartite, undirected) & $\rho(\alpha) = \dfrac{|E_{\alpha}|}{n_A^{\alpha}\,n_B^{\alpha}}$ & $n_A^{\alpha},\,n_B^{\alpha}$: sizes of the two bipartite sets in $\alpha$ \\[10pt]
11 & Density (bipartite, directed) & $\rho(\alpha) = \dfrac{|E_{\alpha}|}{2\,n_A^{\alpha}\,n_B^{\alpha}}$ & \\[8pt]
12 & Average density & $\bar{\rho} = \dfrac{1}{L}\displaystyle\sum_{\alpha=1}^{L} \rho(\alpha)$ & \\
\addlinespace[3pt]
\midrule
\multicolumn{4}{@{}l}{\textit{Pairwise (between layers $\alpha$ and $\beta$)}} \\
\addlinespace[2pt]
13 & Jaccard, node identity & $J^{\mathrm{node}}_{\alpha\beta} = \dfrac{|X_{\alpha}\cap X_{\beta}|}{|X_{\alpha}\cup X_{\beta}|}$ & Undefined when both sets are empty \\[10pt]
14 & Jaccard, edge identity & $J^{\mathrm{edge}}_{\alpha\beta} = \dfrac{|E_{\alpha}\cap E_{\beta}|}{|E_{\alpha}\cup E_{\beta}|}$ & On deduplicated edge keys \\
\addlinespace[3pt]
\midrule
\multicolumn{4}{@{}l}{\textit{Meta-graph aggregation (projection over all layers)}\tnote{a}} \\
\addlinespace[2pt]
15 & Meta-edge weight, union & $w_{\mathrm{meta}}(u,v) = \mathbf{1}\!\left[\exists\,\alpha\,:\,(u,v)\in E_{\alpha}\right]$ & Unweighted union; recovers $\mathrm{proj}(\mathcal{M})$ of \cite{Boccaletti2014-vr} \\[6pt]
16 & Meta-edge weight, sum-occurrence & $w_{\mathrm{meta}}(u,v) = \displaystyle\sum_{\alpha} \mathbf{1}\!\left[(u,v)\in E_{\alpha}\right]$ & Counts layers in which the edge appears \\[6pt]
17 & Meta-edge weight, sum-weights & $w_{\mathrm{meta}}(u,v) = \displaystyle\sum_{\alpha} a^{\alpha}_{uv}$ & Sums layer-specific weights ($a^{\alpha}_{uv}=0$ if absent) \\[6pt]
18 & Meta-degree & $k^{\mathrm{meta}}(v) = \left|\{u\,:\,(u,v)\in E_{\mathrm{meta}}\}\right|$ & Distinct from $k_{\mathrm{inter,sum}}(v)$: operates on the projection, not on cross-layer links \\[6pt]
19 & Meta-strength & $s^{\mathrm{meta}}(v) = \displaystyle\sum_{u:\,(u,v)\in E_{\mathrm{meta}}} w_{\mathrm{meta}}(u,v)$ & Weighted analogue of (18); weights from row 15, 16, or 17 \\
\bottomrule
\end{tabular}
\begin{tablenotes}\footnotesize
\item[a] Edge canonicalisation is applied before aggregation: directed edges keep their order $(u,v)$; undirected edges use the sorted pair as the canonical key, so $(u,v)$ and $(v,u)$ are not double-counted.
\end{tablenotes}
\end{threeparttable}
\end{table}

\subsection*{Visualization complexity and performance}
Multilayer networks are inherently more visually complex than monolayer networks. The three-dimensional stacked representation adds structural depth but also visual density: as layer count and link count increase, the visualization becomes harder to navigate and interpret. MiRA enables users to address this through a combination of complementary visualization modes, interactive filtering, and link-weight thresholding, allowing users to progressively simplify what is rendered to match the goals of their visualization (e.g., focusing on overall structure, node connectivity within and between layers, or node attributes across layers). Rendering performance scales primarily with the number of visible elements. MiRA renders using the browser's Canvas 2D API, which is CPU-bound rather than GPU-accelerated. In practice, interlayer links are the dominant performance cost: they are hidden by default and should be revealed selectively. To establish rendering limits, we stress-tested MiRA with a synthetic temporal network of 8 layers, 100 physical nodes, 8,000 intralayer links, and 7,000 interlayer links — a network far denser than would be interpretable in its entirety. On a MacBook Pro M1 and and on a Windows with Intel I7, both running Google Chrome , all 15,000 links rendered without issues and navigation remained fluid; performance will vary on lower-powered hardware. We note that visualizing such a network in full is not recommended in practice: the analytical value of MiRA lies in progressive exploration — filtering, thresholding, and switching modes — rather than presenting the complete multilayer network at once.

\subsection*{Software architecture and implementation}

MiRA is implemented entirely in JavaScript as a static, single-page web application with no server-side components and no build pipeline. The application loads directly in any modern browser from a single HTML entry point on laptop and desktop computers. Its architecture is not designed for tablets or phones. MiRA's functionality is organized into discrete ES modules. This modular codebase enforces separation of concerns between data ingestion, layout computation, and rendering, and facilitates independent testing and maintenance of each subsystem. External dependencies are JavaScript libraries loaded at runtime: \texttt{Leaflet} for tile-based interactive maps, \texttt{PapaParse} for CSV parsing, \texttt{d3} for force-directed layouts and number formatting, \texttt{Konva} for off-screen hit-testing, and \texttt{chroma.js} for color-scale construction; \texttt{html2canvas} and \texttt{jsPDF} are loaded on demand by the export pipeline. No 3D library is used: the layered visualization is rendered to a Canvas 2D context using a hand-rolled oblique isometric projection.

\subsection*{Deployment and offline use}

The production instance of MiRA is hosted on GitHub Pages and requires no installation. Because the application is entirely static, it can equally be run offline: users may clone the repository and open the \texttt{index.html} file directly in a browser, with no web server required. A copy of the application is also bundled with the EMLN R package \cite{Frydman2023-hg}; when \texttt{plot\_multilayer()} is called, EMLN serializes the network object to JSON, starts a local HTTP server via \texttt{httpuv}, and opens the bundled application in the default browser with the network pre-loaded. This offline pathway ensures reproducibility when the hosted version evolves, as the version of MiRA bundled with a given EMLN release is fixed at installation time. For researchers using R, it also ensures consistency between the analytical and visualization phases of the workflow.

\subsection*{App development}
We developed the MiRA codebase using an AI-assisted workflow. Initial scaffolding and iterative feature implementation were carried out using Claude Code (Anthropic), an agentic AI programming tool, with Claude Sonnet~4.6 used for routine implementation tasks and Claude Opus~4 used for complex architectural decisions and refactoring. Human oversight was applied throughout: all AI-generated code was reviewed, debugged, and revised. We paid particular attention to calculation of properties and to domain-specific logic, including bipartite layout computation, geographic placement of layers, multilayer data structure validation, and EMLN integration. The codebase was audited periodically to identify inconsistencies between behavior and documentation, to enforce modularity via refactoring, and to remove redundancy introduced during iterative development.

\section*{Code availability}
MiRA is open-source under the Creative Commons Attribution--NonCommercial--ShareAlike license (CC BY-NC-SA) and freely accessible at \url{https://mira.ecomplab.com/}. Source code is available at \url{https://github.com/Ecological-Complexity-Lab/MiRA} and this paper describes version 1.0. Full technical documentation and a user manual are available at 
\url{https://mira.ecomplab.com/docs/manual.html}. MiRA is integrated into the EMLN R package \cite{Frydman2023-hg}, which is developed and maintained independently and is available at 
\url{https://emln.ecomplab.com/}.

\section*{Funding}
This work was supported by grants from the Israel Science Foundation (grant 491/25), the Israel Council of Higher Education (Long-Term Research of Urban Ecosystems in Israel), and the Human Frontiers Science Program Organization (grant number RGY0064/2022) to SP.
\printbibliography

\end{document}